\documentclass[conference,letterpaper]{IEEEtran}
\usepackage[utf8]{inputenc}
\usepackage[T1]{fontenc}
\usepackage{url}              
\usepackage{cite}             
\usepackage[cmex10]{amsmath}
\usepackage{amssymb}
\usepackage{amsmath}
\usepackage{xifthen}
\usepackage{xcolor}

\usepackage{mleftright} 
\mleftright  

\usepackage{graphicx}
\usepackage[font=footnotesize]{caption}
\usepackage[font=footnotesize]{subcaption}
\usepackage{tabularx,booktabs}
\newcolumntype{C}{>{\centering\arraybackslash}X} 
\usepackage{algorithmicx}        
\usepackage{algorithm}          
\usepackage{algpseudocode}

\usepackage{stfloats}

\usepackage{amsthm}
\usepackage{mathtools}
\usepackage{multirow}
\usepackage{lipsum}  

\newtheorem{theorem}{\bfseries Theorem}
\newtheorem{lemma}{\bfseries Lemma}
\newtheorem{definition}{\bfseries Definition}
\newtheorem{corollary}{\bfseries Corollary}

\newtheorem{remark}{\bfseries Remark}

\newtheorem*{assumptions*}{\bfseries Assumptions}

\newtheorem{problem}{\bfseries Problem}

\DeclareMathOperator{\KL}{D_{KL}}

\DeclareMathOperator*{\argmin}{arg\,min}
\newcommand{\E}[2][]{ \mathbb{E}_{#1} \left[ #2 \right] }
\newcommand{\cdf}[1]{ F_{#1} }
\newcommand{\vcdf}[1]{ \Phi_{#1} }
\newcommand{\icdf}[1]{ Q_{#1}}
\newcommand{\vicdf}[1]{ \Psi_{#1} }
\newcommand{\pdf}[1]{ f_{#1} }
\newcommand{\marg}[1]{ m_{#1} }
\newcommand{\cu}[1]{ c_{#1}}
\newcommand{\Cu}[1]{ C_{#1}}
\newcommand{\bx}{\mathbf{x}}
\newcommand{\by}{\mathbf{y}}
\newcommand{\bu}{\mathbf{u}}

\newcommand{\hQ}{\hat{Q}}
\newcommand{\UI}{[0,1]}
\newcommand{\R}{\mathbb{R}}
\newcommand{\Rc}{\mathbb{\bar{R}}}
\newcommand{\N}{\mathbb{N}}
\newcommand{\EOT}{D_{EOT}}
\newcommand{\Rpr}{R_{PR}}
\newcommand{\Roc}{R_{OC}}
\newcommand{\Rslb}{R_{PR}^{SLB}}
\newcommand{\var}[1]{\sigma^2_{#1}}

\newcommand{\FS}[1]{#1}
\newcommand{\todo}[1]{#1}
\newcommand{\done}[1]{#1}

\begin{document}

\title{Copula-based Estimation of Continuous Sources for a Class of Constrained Rate-Distortion-Functions}

\author{
   \IEEEauthorblockN{Giuseppe Serra,
                    Photios A. Stavrou,
                    Marios Kountouris}
  \IEEEauthorblockA{Communication Systems Department, EURECOM,                    Sophia-Antipolis, France}
\IEEEauthorblockA{\texttt{\{giuseppe.serra,fotios.stavrou,marios.kountouris\}}@eurecom.fr}
}

\maketitle

\begin{abstract} 
We present a {new} method to estimate the rate-distortion-perception function in the perfect realism regime (PR-RDPF), for multivariate continuous sources subject to a single-letter average distortion constraint. The proposed approach is not only able to solve the specific problem but also two related problems: the entropic optimal transport (EOT) and the output-constrained rate-distortion function (OC-RDF), of which the PR-RDPF represents a special case. Using copula distributions, we show that the OC-RDF can be cast as an $I$-projection problem on a convex set, based on which we develop a parametric solution of the optimal projection proving that its parameters can be estimated, up to an arbitrary precision, via the solution of a convex program. Subsequently, we propose an iterative scheme via gradient methods to estimate the convex program. Lastly, we characterize a Shannon lower bound (SLB) for the PR-RDPF under a mean squared error (MSE) distortion constraint. We support our theoretical findings with numerical examples by assessing the estimation performance of our iterative scheme using the PR-RDPF with the obtained SLB for various sources.
\end{abstract}

\section{Introduction}

Rate-distortion-perception (RDP) theory, which provides a way to reconstruct complex data sources (e.g., audio, images, video) when perceptual quality is taken into account in addition to the distortion criterion, has recently attracted increasing interest within the information theory, computer vision, and machine learning communities. This framework, proposed by Blau and Michaeli \cite{blau:2019} and Matsumoto \cite{matsumoto:2018,matsumoto:2019}, generalizes the classical rate-distortion function (RDF) formulation by imposing a divergence constraint between the source distribution and its reconstruction. In RDP theory, the divergence constraint acts as a proxy for human perception, capturing the difference between the reconstructed samples and the source ''natural statistic'' \cite{mittal:2013}. It can also be used as a semantic quality metric measuring the relevance of the reconstructed source from the receiver's perspective \cite{kountouris:2020}. 
\par Prior to the development of the RDP theory, a similar line of research in lossy compression has studied the link between the statistical properties of the distribution of the reconstructed samples and their perceptual quality, defining the so-called \textit{output-constrained rate-distortion problem} \cite{li:2011:distribution,saldi:2015,saldi:2015:2}. In this class of constrained lossy compression problems, instead of restricting the maximal statistical divergence between the source distribution and its reconstruction, the focus is on constraining the reconstruction to belong to a specific distribution, which may differ from that of the source. The resulting problem is in close proximity to the  EOT problem \cite{bai:2020, wang:2022}. Interestingly, in both problems, the source and the reconstruction distributions are assumed to be known {\it a priori}. 
\par The mathematical formulation that quantifies the operational meaning in RDP theory is the RDPF, which, much like its classical RDF counterpart, is not generally available in analytical form. Despite the general complexity, closed-form expressions have been developed under different settings \cite{blau:2019,zhang:2021, serra:2023:computation, qian:2023}. The absence of a general analytic solution for the RDPF led to the research of computational methods for its estimation. However, dedicated algorithmic solutions have been developed so far only for discrete sources \cite{serra:2023} or by discretizing certain classes of continuous sources \cite{chen:2023}. For general sources, RDPF estimation methods often rely on data-driven solutions \cite{blau:2019, zhang:2021, ogun:2021}, which unfortunately do not have convergence guarantees.

\subsection{Contributions}
In this work, we propose a new copula-based estimation method for the computation of the PR-RDPF for multivariate continuous sources subject to a single-letter average distortion constraint. Our estimation method is quite general as it also allows the computation of the EOT and the OC-RDF for which the PR-RDPF is a particular case. 
\par The main contributions of this paper are as follows. {\bf (i)} We show that there exists a one-to-one correspondence between the feasible set of solutions of the OC-RDF and EOT (Theorem \ref{theorem:EquivalenceEOTRD}), making the two problems equivalent. {\bf (ii)} Using properties of copula distributions, we demonstrate that the OC-RDF can be reformulated as a projection problem in the geometry induced by the Kullback–Leibler (KL)-divergence, i.e., $I$-projection, on a convex constraint set (Problem \ref{problem:ProjectionInfoRDPF}). However, although this class of projection has been extensively studied in \cite{Csiszar:1975}, the existing parametric solution is not directly suitable for computational purposes. To bypass this technical issue, we introduce a relaxation of the constraint set of the $I$-projection, which results in a lower bound to the original optimization objective (Problem \ref{problem:MomentConstrainedRDPF}) that we subsequently show that it can be made arbitrarily tight (Theorem \ref{theorem:convergenceRelaxation}). {\bf (iii)} We characterize the parametric closed-form solution of the relaxed $I$-projection, whose optimal parameters can be directly obtained as the solution of a strictly convex program (Theorem \ref{theorem:ELProjection}). 
{\bf (iv)} We propose an algorithmic approach via a stochastic gradient descent method, to estimate the strictly convex optimization problem of Theorem \ref{theorem:ELProjection} (see Alg. \ref{alg:MIminimization}). {\bf (v)} We derive a Shannon lower bound (SLB) for the PR-RDPF under MSE distortion (Theorem \ref{th:ShannonLowerbound}). 
\par We supplement our theoretical results with various numerical evaluations aiming to estimate the PR-RDPF under various sources and different distortion measures via Alg. \ref{alg:MIminimization}, and to demonstrate the efficacy of our algorithmic approach compared to the obtained SLB.

\begin{subsection}{Notation}
   Given a Polish space $\mathcal{X}$, we denote by $(\mathcal{X}, \mathbb{B}(\mathcal{X}))$ the Borel measurable space induced by the metric, with $\mathcal{P}(\mathcal{X})$ denoting the set of distribution functions defined thereon. For a random variable (RV) $X$ defined on $(\mathcal{X}, \mathbb{B}(\mathcal{X}))$, we denote with $\cdf{X} \in \mathcal{P}(\mathcal{X})$ its distribution function (shortly, d.f.) and with $\pdf{X}$ its probability density function (shortly, pdf). Given two RVs $X$ and $Y$, we will indicate their independent product d.f. as $\cdf{X} \otimes \cdf{Y}$, equivalent to the independent product pdf $\pdf{X,Y} = \pdf{X}\pdf{Y}$. Furthermore, given any joint pdf $\pdf{X,Y}$, we will indicate with $\marg{X}(\pdf{X,Y})$ and $\marg{Y}(\pdf{X,Y})$ the pdf associated with the marginal RV's $X$ and $Y$, {respectively}. \done{We will indicate with $\KL(\cdf{X}||\cdf{Y})$ the Kullback–Leibler (KL)-divergence between RV's $X$ and $Y$, whereas $h(X)$ and $h(X|Y)$ will denote, respectively, the differential entropy of $X$ and the conditional differential entropy of $X$ given $Y$. We indicate with $\R$ the set of real numbers, with $\Rc$ the extended set $\R \cup \{-\infty, +\infty \}$. Lastly, given a set $\mathcal{A} \in \R^n$, we  will denote with $l_p(\mathcal{A})$ the set of functions $g:\mathcal{A} \to \R$ such that $\int_{\mathcal{A}} |g(s)|^pds < \infty$}.
\end{subsection}

\section{Preliminaries}

\subsection{OC-RDF - A link between PR-RDPF and EOT}

We begin this section by providing the {mathematical definition of PR-RDPF}.

\begin{definition}{(PR-RDPF)}\label{problem:RDPF}
Let $\pdf{X} \in \mathcal{P}(\mathcal{X})$. Then, the PR-RDPF for the source {$X \sim \pdf{X}$} under a distortion measure $\Delta:\mathcal{X} \times \mathcal{Y} \to \mathbb{R}^+_0$ is {given as follows} 
{
\begin{align*}
        \Rpr(D) & =  \min_{\substack{f_{Y|X}\\  
        \E{\Delta(X,Y)} \le D\\
        X \sim Y}}  I(X, Y)
\end{align*}
}
where the minimization is on set of Markov kernels $\pdf{Y|X}$.
\end{definition}

{It should be noted that the \textit{perfect realism}} regime represents a limit case of the general problem of the  RDPF \cite{blau:2019}, where one constrains the reconstruction $Y$ to have the same distribution as the source $X$. {Although PR-RDPF became quite popular through \cite{blau:2019}}, similar ideas were previously explored by Li \textit{et. al.} in \cite{li:2011:distribution}, {in the context of} distribution-preserving quantization and distribution-preserving RDF. Multiple coding theorems have been developed for {PR-RDPF}. {For instance,} Chen {\it et. al.} in \cite{chen:2022} proves the necessity of some form of randomness, either private or common, between {the} encoder and decoder, to achieve {the} {\it perfect realism} regime and derives the associated coding theorems.  Wagner, in \cite{wagner:2022:rate}, provides a coding theorem for the RDPF trade-offs for the perfect and near-perfect realism cases, when only finite common randomness between the encoder and decoder is available.
\par {Although our primary goal in this work is to study computational aspects of the PR-RDPF for continuous sources, we do it by also studying a generalization of this problem. {In particular,} we  study the problem of OC-RDF that was formally introduced by Saldi {\it et al.} in \cite{saldi:2015} (see also \cite{li:2011:distribution}), for which the mathematical definition is stated next.}

\begin{definition}(OC-RDF)\label{problem:OC-RDF}
Let $\pdf{X} \in \mathcal{P}(\mathcal{X})$. Then, the {OC-RDF} for the source {$X \sim \pdf{X}$} under a distortion measure $\Delta:\mathcal{X} \times \mathcal{Y} \to \mathbb{R}^+_0$ and a target reconstruction distribution $\pdf{Y} \in \mathcal{P}(\mathcal{Y})$ is {given as follows}
{
\begin{align}
        \Roc(D) = & \min_{\substack{f_{Y|X} \in \hat{\Pi}(\pdf{X}, \pdf{Y})\\ \E{\Delta(X,Y)} \le D}}  I(X, Y) \label{opt:RDPF} 
\end{align}
}
where the minimization is on the convex set of Markov kernels $\hat{\Pi}(\pdf{X}, \pdf{Y}) \triangleq \{ \pdf{X|Y}: \marg{Y}(\pdf{Y|X} \cdot \pdf{X}) = \pdf{Y} \}$.
\end{definition}

The main difference between {the problems of} PR-RDPF and OC-RDF lies in how the constraint on the reconstruction distribution $\pdf{Y}$ is handled. While in the PR-RDPF case, we specifically constrain the reconstruction distribution and source distribution to be identical, {in the} OC-RDF {we have an additional degree of freedom}, allowing for the distribution of the reconstruction to be chosen freely. {This results in the following observation.}
\begin{remark}
{The problem of the OC-RDF particularizes to the problem of PR-RDPF} by specifying the reconstruction distribution to be equal to the source distribution (i.e. $\pdf{Y} = \pdf{X}$).
\end{remark}

{Additionally, the OC-RDF highlights an interesting connection to the EOT problem (see \cite{bai:2020, wang:2022}), of which the mathematical definition is stated as follows.} 

\begin{definition} \label{problem:EOT}
{(EOT)} Let $\pdf{X}\in \mathcal{P}(\mathcal{X})$ and $f_{Y} \in \mathcal{P}(\mathcal{Y})$. Then, {the EOT} for $\epsilon > 0$ and distortion measure $\Delta:\mathcal{X} \times \mathcal{Y} \to \mathbb{R}^+_0$, is {given as follows}
\begin{align}
    \EOT(\epsilon) = \min_{\pdf{X,Y} \in \bar{\Pi}(\pdf{X},\pdf{Y})} \E{\Delta(X,Y)} + \epsilon I(X,Y)
\end{align}
where the minimization is on the convex set of joint pdfs $\bar{\Pi}(\pdf{X}, \pdf{Y}) \triangleq \{ \pdf{X,Y}: \marg{X}{(\pdf{X, Y})} = \pdf{X}, \marg{Y}{(\pdf{X, Y})} = \pdf{Y}\}$.
\end{definition}

{Notably}, it can be shown that {OC-RDF and EOT  are closely related in the sense that for specific values of $D$ and $\epsilon$, there exists a one-to-one mapping between the sets of solutions of the two problems. In other words, we can find the solution to one problem based on the solution of the other}. {To the best of our knowledge, this observation has not been previously documented elsewhere, hence we formalized it in the following theorem.} 

\begin{theorem}{(Connection of OC-RDF and EOT)}\label{theorem:EquivalenceEOTRD}
    Let $\pdf{X} \in \mathcal{P}(\mathcal{X})$ and $\pdf{Y} \in \mathcal{P}(\mathcal{Y})$. Then, for any $D > 0$, there exists {an $\epsilon>0$} such that {the problems of OE-RDF and EOT} are equivalent.
\end{theorem}
\begin{IEEEproof}
    See Appendix \ref{proof:EquivalenceEOTRD}.
\end{IEEEproof}

{In view of} Theorem \ref{theorem:EquivalenceEOTRD}, {we can} treat the OC-RDF and EOT problems as equivalent problems. {As a result,} the computational schemes derived in Section \ref{sec:MainResults} {applicable to} the OC-RDF problem, can be adapted {\it mutatis mutandis} to the EOT problem. 

\subsection{Copula distributions} \label{sec:WhatIsACopula?}
{In this subsection, we give some preliminaries to copulas distributions, as these have a central role in the derivation of the main results of this paper. The following definitions and theorems are taken from \done{\cite{durante:2010}}.} 

\begin{definition}{\it (Copula distribution)} For every $d \ge 2$, a $d$-dimensional copula d.f. is a $d$-variate d.f. on $\UI^d$ whose univariate marginals are uniformly distributed on $\UI$.
\end{definition}

\par {The next theorem and the two companion corollaries, demonstrate that copulas are a powerful tool for the modeling and analysis of multivariate distributions.} 

\begin{theorem}({\it Sklar's Theorem}) Let F be a $d$-dimensional d.f. with marginal d.f. $F_1, F_2, \ldots, F_d$. Let $A_j$ denote the range of $F_j$, $A_j  \triangleq F_j \left( \Rc \right) \quad (j = 1,2, \ldots, d)$. Then, there exists a $d$-copula d.f. $C$ such that for all $(x_1,x_2, \ldots, x_d) \in \Rc^d$,
\begin{align}
    F(x_1, \ldots, x_d) = C\left(F_1(x_1), \ldots, F_d(x_d)\right). \label{definition:copula:Sklard}
\end{align}
Such a $C$ is uniquely determined on $A_1 \times A_2 \times \cdots A_d$ and, hence, it is unique when $F_1, F_2, \ldots, F_d$ are continuous.
\end{theorem}
\begin{corollary} \label{corollary:SklarTheoremForPdf}
    Let $f:\Rc^d \to \R^+$ be the pdf associated with \eqref{definition:copula:Sklard}. Then, $f$ can be uniquely decomposed as
    \begin{align}
        f(x_1,\ldots, x_d) = c\left(F_1(x_1), \ldots, F_d(x_d)\right) \prod_{j=1}^d f_j(x_j)
    \end{align}
    where $f_j$ is the pdf associated with the univariate marginal d.f. $F_j$ and $c:\UI^d \to \R^+$ is the pdf associated with the copula d.f. $C$. 
\end{corollary}
\begin{corollary} \label{corollary:CopulaConstruction}
    Let $F_1, F_2, \ldots, F_d$ be univariate d.f.'s and $C$ be a copula d.f.. {Then}, the function $F:\Rc^d \to \UI$ defined in \eqref{definition:copula:Sklard} is a d-dimensional d.f. with marginal $F_1, F_2, \ldots, F_d$.
\end{corollary}
It is worth noticing that Corollary \ref{corollary:SklarTheoremForPdf} guarantees that the pdf of any multivariate distribution can be factorized as the product of the marginal densities and a unique copula distribution. This factorization can be effectively thought of as decoupling the correlation structure embedded in the joint distribution (represented by the copula distribution) from the information regarding each single marginal. {On the other hand,} Corollary \ref{corollary:CopulaConstruction} guarantees that, {for a fixed} set of marginals distributions, any copula distribution describes a proper joint distribution.  
\par We conclude this {subsection} with the definition of the quantile function, which will {also be of use in the derivation of our main results}.

\begin{definition}\textit{(Quantile function)} {Let} $X \sim F_X$ be a univariate {RV} on $\mathcal{X} \subseteq \R$. We define the quantile function $\icdf{X}:\UI \to \R$ as $\icdf{X}(u) \triangleq \sup \{ x \in \mathcal{X}: F(x) \le u \}$. If $\cdf{X}$ is continuous and strictly increasing, then $\icdf{X} = F^{-1}_X$. However, even if $\cdf{X}$ may fail to have an inverse function, $\icdf{X}$ guaranties that $\icdf{X}\left( \cdf{X}(X) \right) = X$ almost surely (a.s.).    
\end{definition}

To ease the notation, in the sequel we denote by \textbf{uniform transformation} of an RV $X = (X_1,\ldots,X_d)$ the function $\vcdf{X}:\mathcal{X} \to \UI^d$ defined as $\vcdf{X}(X) \triangleq (\cdf{X_1}(X_1), \dots, \cdf{X_d}(X_d))$. Moreover, we define the function $\vicdf{X}:\UI^d \to \mathcal{X}$ as $\vicdf{X}(U) \triangleq (\icdf{X_1}(U_1), \dots, \icdf{X_d}(U_d))$.  By construction, $\vicdf{X}$ is the a.s.-inverse of $\vcdf{X}$, that is, $\vicdf{X}(\vcdf{X}(X)) = X$ a.s.

\section{Main Results} \label{sec:MainResults}

{In this section, we derive our main results.}

\subsection{Copula Lower Bound}
{First, we prove a lemma with which the functionals in the mathematical formulations of Definitions \ref{problem:OC-RDF} and \ref{problem:EOT} can be redefined using copula distributions.}

\begin{lemma} \label{lemma:MI2KL}
Let $(X,Y) \sim \pdf{XY} \in \mathcal{P}(\mathcal{X} \times \mathcal{Y})$ be a $2d$-variate {RV} with marginal pdfs $\pdf{X} \in \mathcal{P}(\mathcal{X})$ and $\pdf{Y} \in \mathcal{P}(\mathcal{Y})$. Then, the mutual information $I(X,Y)$ {can be equivalently written as follows} 
\begin{align}
    I(X,Y) = \KL(C_{X,Y}||C_X \otimes C_Y) \label{eq:copula:generalMI}
\end{align}
where $C_{X,Y}, C_X, C_Y$ are the copula d.f.'s associated with distributions $\cdf{X,Y}$, $\cdf{X}$, and $\cdf{Y}$, {respectively}. {In addition}, given a distortion function $\Delta:\mathcal{X} \times \mathcal{Y} \to \R^+$, the following holds
\begin{align}
    \E[\cdf{X,Y}]{\Delta(X,Y)} = \E[\Cu{X,Y}]{\Delta\left(\vicdf{X}(U_X),\vicdf{Y}(U_Y)\right)} \label{eq:copula:distortion}
\end{align}
where $U = (U_X,U_Y) \sim \Cu{X,Y}$.
\end{lemma}
\begin{IEEEproof}
    See Appendix \ref{proof:MI2KL}. 
\end{IEEEproof}

{Leveraging Lemma \ref{lemma:MI2KL}, we can provide an alternative formulation of the mathematical expression in \eqref{opt:RDPF}, which will be the subject of our estimation analysis. This is stated next as Problem \ref{problem:ProjectionInfoRDPF}.} 

\begin{problem}{(Copula-based OC-RDF)} \label{problem:ProjectionInfoRDPF}
{The mathematical expression \eqref{opt:RDPF} can be reformulated as follows}
    \begin{align}
        \Roc(D) &=  \min_{\Cu{} \in \mathcal{\Cu{}}_{2d}}  \KL(\Cu{}||{\Cu{X} \otimes \Cu{Y}}) \label{opt:CopulaKL} \\
        \textrm{s.t.}   & ~\E[C]{\Delta(\vicdf{X}(U_X),\vicdf{X}(U_Y))} = D \label{opt:CopulaDistor}
    \end{align}
where $\mathcal{\Cu{}}_{2d}$ is the set of $2d$-copula distributions and $D \in [D_{\min}, D_{\max}]$. 
\end{problem}

\begin{remark} (On Problem \ref{problem:ProjectionInfoRDPF})
 Problem \ref{problem:ProjectionInfoRDPF} is a convex program in the space of copula d.f. {Moreover,} the problem is equivalent to finding the $I$-projection of {$\Cu{X} \otimes \Cu{Y}$} on the set $\mathcal{B} \subset \mathcal{C}_{2d}$ of copula d.f. satisfying the modified distortion constraint \eqref{opt:CopulaDistor}.
\end{remark}

Problem \ref{problem:ProjectionInfoRDPF} represents a {projection} problem in {information geometry}, where the goal is to find the copula distribution $C$ that minimizes the information divergence from the independent product copula $\Cu{X} \otimes \Cu{Y}$ while respecting a linear set of constraints. This class of {projection problems} has been {thoroughly} studied by Csisz\'ar in \cite{Csiszar:1975}, where the analytical form of the optimal projection for the considered case has been characterized. {Using \cite{Csiszar:1975}, we derive the following theorem.}

\begin{theorem}{(Analytical solution of Problem \ref{problem:ProjectionInfoRDPF})} \label{theorem:OptimalProjection}
Let $R = \Cu{X} \otimes \Cu{Y} $ and assume there exists a copula d.f. $P$  such that ${\KL(P||R)} < \infty$ and \eqref{opt:CopulaDistor} is satisfied. Then, Problem \ref{problem:ProjectionInfoRDPF} admits {a} minimizing copula $Q$ with Radon–Nikodym derivative with respect to the measure $R$ of the form 
\begin{align}
\frac{d\Cu{}}{dR}(\bu) = e^{\mu + \theta[\Delta(\vicdf{X}(\bu_x),\vicdf{Y}(\bu_y))]} \prod_{i = 1}^{2d} g_{i}(u_{i}) \label{eq:copula:optimaldensity}
\end{align}
for some constants $(\mu, \theta)$, and nonnegative uni-variate functions $g_i$ such that $\log(g_i(s)) \in l_1(\UI)$ for $i=1,\ldots,2d$.
\end{theorem}
\begin{IEEEproof}
    See Appendix \ref{proof:OptimalProjection}.
\end{IEEEproof}

{Although} Theorem \ref{theorem:OptimalProjection} provides a characterization of the solution of Problem \ref{problem:ProjectionInfoRDPF}, the lack of an analytical form for the free functions $\{g_i(\cdot)\}_{i=1\ldots,2d}$ poses a challenging problem in the computation of \eqref{eq:copula:optimaldensity}.
{Following an idea of} \cite{samo:2021:copula}, we circumvent this technical {issue} by introducing a relaxation on the constraint set {of} Problem \ref{problem:ProjectionInfoRDPF}, {that results into} a lower bound on OC-RDF. {This is demonstrated next in Problem \ref{problem:MomentConstrainedRDPF}.} 

\begin{problem}(Lower bound to {Problem~ \ref{problem:ProjectionInfoRDPF}}) \label{problem:MomentConstrainedRDPF} \done{For any integer $N$}, Problem \ref{problem:ProjectionInfoRDPF} can be lower bounded {as follows}
    \begin{align*}
        \Roc(D) \ge \Roc^{(N)} =& \min_{\substack{Q \in \mathcal{D}(\UI^{2d})\\ \E{\Delta(\vicdf{X}(U_X),\vicdf{Y}(U_Y))} = D\\
        \E[Q]{u_{i}^n} = \alpha_n,~~(i,n) \in I}}  \KL(Q|| R )
    \end{align*}
where $R = \Cu{X} \otimes \Cu{Y}$, $I = (1,\ldots,2d) \times (1,\ldots,N)$, $D \in [D_{\min}, D_{\max}]$, and $\alpha_n$ is the $n^{th}$ moment of a uniform distribution on $\UI$. 
\end{problem}

\begin{remark} (Problem \ref{problem:ProjectionInfoRDPF} vs Problem \ref{problem:MomentConstrainedRDPF})
The main {technical} difference between Problems \ref{problem:ProjectionInfoRDPF} and \ref{problem:MomentConstrainedRDPF} concerns their constraint sets.
{Particularly, in Problem \ref{problem:ProjectionInfoRDPF} we require that the minimizing distribution $Q^*$ belongs to the set of copula distributions, which means that its marginals are uniformly distributed. On the other hand, the marginals of the minimizing distribution $\hQ^*_N$ of Problem \ref{problem:MomentConstrainedRDPF} only require to respect up to $N$ moments of a uniform distribution. This in turn} implies that the constraint set of Problem \ref{problem:ProjectionInfoRDPF} is a proper subset of the constraint set of Problem \ref{problem:MomentConstrainedRDPF}, justifying the lower bound {of the latter}. 
\end{remark}

{In the following theorem, we show that, for $N \to \infty$, Problem \ref{problem:MomentConstrainedRDPF} recovers the solution of Problem \ref{problem:ProjectionInfoRDPF}.}
\begin{theorem} \label{theorem:convergenceRelaxation}
    Let $Q^*$ be the optimal solution of Problem \ref{problem:ProjectionInfoRDPF} and $\hQ^*_N$ be the optimal solution of Problem \ref{problem:MomentConstrainedRDPF}. Then, as ${N \to \infty},$
    \begin{align*}
         \KL(\hQ^*_N || Q^*) \to 0  ~\text{and}~ \Roc^{(N)} \to \Roc.
    \end{align*}
\end{theorem}
\begin{IEEEproof}
\todo{See  Appendix  \ref{proof:convergenceRelaxation}.}
\end{IEEEproof}

We now provide the analytical form of the solution of Problem \ref{problem:MomentConstrainedRDPF}. Unlike Theorem \ref{theorem:OptimalProjection}, the optimal solution does not depend on free functions $\{g_i(\cdot)\}_{i=1\ldots,2d}$, but it depends only on the Lagrangian multipliers of Problem \ref{problem:MomentConstrainedRDPF} {obtained} as result of its dual problem. 

\begin{theorem}{(Analytical solution of Problem \ref{problem:MomentConstrainedRDPF})} \label{theorem:ELProjection}
    Let $R = \Cu{X} \otimes \Cu{Y} $ and assume there exists a d.f. $P$ on $\UI^{2d}$ such that ${\KL(P||R)} < \infty$ and \eqref{opt:CopulaDistor} is satisfied. Then, Problem \ref{problem:MomentConstrainedRDPF} admits minimizing copula $Q$ with Radon–Nikodym derivative with respect to the measure $R$ of the form 
    \begin{align}
        \frac{dQ}{dR}(\bu) = e^{\mu + \theta\Delta(\vicdf{X}(\bu_x),\vicdf{Y}(\bu_y))} \prod_{i = 1}^{2d} e^{\sum_{n=0}^N \nu_{i,n} u_{i}^n} \label{eq:copula:ELdensity}
    \end{align}
    where the constants $(\mu, \theta, \{ \nu_{i,n} \}_{(i,n)\in I} )$ are the Lagrangian multipliers of Problem \ref{problem:MomentConstrainedRDPF} {obtained as a result of} the following dual program
    \begin{align}
        \begin{split}
        \min_{(\mu, \theta, \{ \nu_{i,n} \}_{(i,n)\in I})} &-\mu -\theta D -\sum_{(i,n) \in I}  \nu_{i,n} \alpha_n \\
        &+ \left(\int_{\UI^{2d}} \frac{dQ}{dR}(\bu) dR(\bu) - 1 \right). 
        \end{split}\label{eq:copula:ELdual}
    \end{align}
\end{theorem}
\begin{IEEEproof}
    See  Appendix \ref{proof:ELProjection}.
\end{IEEEproof}
{The following result is a consequence of Theorem \ref{theorem:ELProjection}.}
\begin{corollary}
Let $Q$ be the {minimizing} copula d.f. characterized in Theorem \ref{theorem:ELProjection}. Then, the mutual information $I(X,Y)$ of the joint distribution $(X,Y)$ defined by marginals {d.f.} $\{\cdf{X_i}\}_{i = 1,\ldots,d}$ and $\{\cdf{Y_i}\}_{i = 1,\ldots,d}$ and copula $Q$ is given by
\begin{align}
I(X,Y) = \KL(Q||R) = -\mu -\theta D -\sum_{(i,n) \in I}  \nu_{i,n} \alpha_n \label{eq:OptimalMI}.
\end{align}
\end{corollary}

\subsection{Copula Estimation}
As anticipated in Theorem \ref{theorem:ELProjection}, the Lagrangian multipliers $(\mu, \theta, \{ \nu_{i,n} \}_{(i,n)\in I})$ defining the optimal solution of Problem \ref{problem:MomentConstrainedRDPF} can be obtained by solving \eqref{eq:copula:ELdual}. Although not available in closed form, the solution of \eqref{eq:copula:ELdual} can be optimally computed using numerical methods, given the properties of the problem.

\begin{lemma} \label{lemma:StrictCovexity}
The optimization problem \eqref{eq:copula:ELdual} is strictly convex, hence it has a unique solution.
\end{lemma}
\begin{IEEEproof}
    See  Appendix  \ref{proof:StrictConvexity}.
\end{IEEEproof}

{To compute \eqref{eq:copula:ELdual}}, we propose a low-complexity optimization scheme based on gradient methods. 
\todo{The main technical detail to clarify is related to the estimation of the integral present in \eqref{eq:copula:ELdual}, since numerically solving a possibly high dimensional integral could hinder the complexity of the algorithm. However, since its computation is required only for the estimation of the gradient and not for the computation of $I(X,Y)$ (as shown in \eqref{eq:OptimalMI}), we can approximate the integral using Monte Carlo method \cite{robert:1999monte}}. 
The resulting iterative scheme can be considered as a {\it mini-batch stochastic gradient descent algorithm} on a convex objective \cite{garrigos2023handbook}. The algorithm is given in Alg. \ref{alg:MIminimization}.

\begin{algorithm}
    \caption{$\Roc(D)$ - Copula Estimation} \label{alg:MIminimization}
    \begin{algorithmic}[1]
        
\Require marginal distributions $\{\cdf{X_i}, \cdf{Y_i}\}_{i = 1,\ldots,d}$; distortion level $D$; number of iterations $T$; initial Lagrangian multipliers $\mathbf{l^{(0)}} = (\mu^{(0)}, \theta^{(0)}, \{ \nu^{(0)}_{i,n} \}_{(i,n)\in I})$; 
        
        \For i = 1, \dots, $T$
            \State Sample $\{\mathbf{u}_i\}_{i = 1\ldots M}$ with  $u_i \sim U(\UI^{2d})$
            \State $f(\mathbf{l}) \approx  \eqref{eq:OptimalMI} + \left(\tfrac{1}{M} \sum_{i = 1}^M \frac{dQ}{dR}(\mathbf{l}, \bu_i) dR(\bu_i) \right)$
            \State $\mathbf{l^{(i)}} = \texttt{GradientMethod}(\mathbf{l}^{(i-1)}, f)$ 
        \EndFor

        \Ensure \done{Lagrangian multipliers $\mathbf{l}^{(T)}$; $I(X,Y) = \eqref{eq:OptimalMI}$}.
    \end{algorithmic}
\end{algorithm}

\subsection{SLB for PR-RDPF} \label{sec:ShannonLowerbound}

{In this {subsection}, we prove a generalization of the well-known SLB on the classical RDF with MSE distortion \cite{berger:1971} to the case of PR-RDPF, denoted hereinafter by $\Rslb$.} The bound is stated in the following theorem. 
\begin{theorem} \label{th:ShannonLowerbound} (SLB for PR-RDPF) Let $\mathcal{S} \triangleq \{\pdf{X}: \E[\pdf{X}]{(X - \E{X})(X -\E{X})^T} \preceq \Sigma \}$ be the set of source distribution {with} a fixed covariance matrix $\Sigma$. Then, for all $X \sim \mathcal{S}$, the PR-RDPF under MSE distortion constraint {admits the following lower bound} 
\begin{align}
\Rpr(D) \ge \Rslb(D) = h(X) - h(X^*) + \Rpr^G(D) \label{eq:shannonLB}
\end{align}
where $\Rpr^G(D)$ denotes the Gaussian PR-RDPF for a source $X^* \sim N(0, \Sigma)$. 
\end{theorem}
\begin{IEEEproof}
\todo{See  Appendix  \ref{proof:ShannonLowerbound}.}
\end{IEEEproof}
{We stress the following technical remark on Theorem \ref{th:ShannonLowerbound}.}

\begin{remark}{(On Theorem \ref{th:ShannonLowerbound})}
    For the scalar case of the {PR-RDPF}, let $\mathcal{S} \triangleq \{\pdf{X}: \E[\pdf{X}]{(X -\E{X})^2} \le \var{}] \}$ for a finite variance value $\var{}$. Then, \eqref{eq:shannonLB} can be further simplified to
\begin{align*}
\Rpr(D) \ge \Rslb(D)=\frac{1}{2} \log \left(\frac{N(X)}{D - \frac{D^2}{4\var{}}} \right)
\end{align*}
with $N(X)$ {denoting} the entropy power of source $X$. For the general vector case, the lower bound depends on the vector Gaussian PR-RDPF, $\Rpr^G$, which can be easily computed using the adaptive reverse-water-filling solution developed in \cite[Corollary 3]{serra:2023:computation}. 
\end{remark}

\section{Numerical Results} \label{sec:NumericalResults}

In this section, we provide numerical estimation of the PR-RDPF for both scalar and vector sources using Alg. \ref{alg:MIminimization}.

\subsubsection*{Scalar Case} 
\begin{figure}
    \centering
    \begin{subfigure}{0.49\linewidth}
    	\centering \includegraphics[width= 0.9\linewidth]{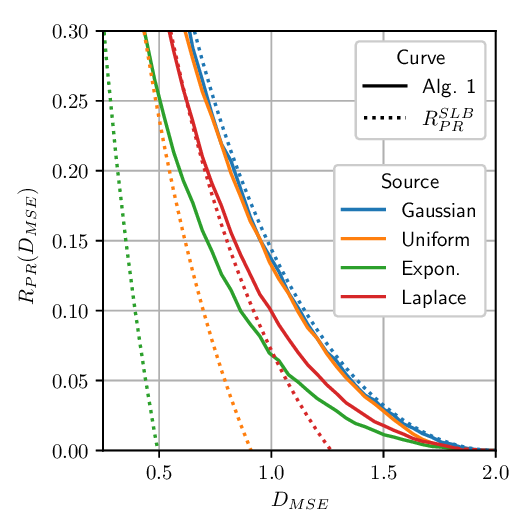}
    	\caption{} \label{fig:RDPF:scalar:L2}
    \end{subfigure}
    \begin{subfigure}{0.49\linewidth}
        \centering \includegraphics[width= 0.9\linewidth]{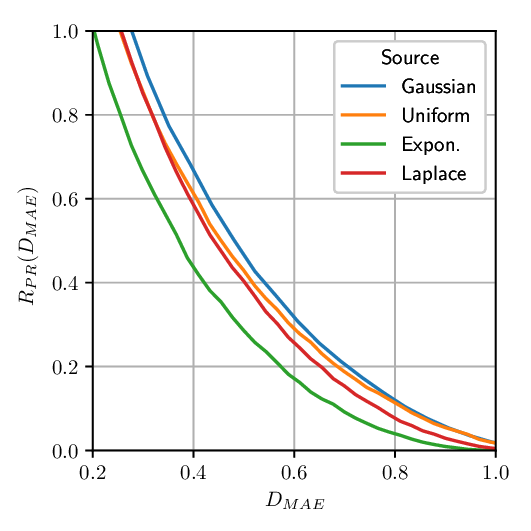}
    	\caption{} \label{fig:RDPF:scalar:L1}
    \end{subfigure}
    \caption{PR-RDPF for various source distributions under (a) MSE distortion metric and (b) MAE distortion metric.} 
\label{fig:scalarCase:RDPF}
\end{figure}

We estimate the PR-RDPF for scalar sources under a single-letter constraint on the reconstruction error in terms of (a) the $l_2$ norm, i.e., the MSE distortion  (see Fig. \ref{fig:RDPF:scalar:L2}), and (b) the $l_1$ norm i.e. the mean-absolute-error (MAE) distortion (see Fig. \ref{fig:RDPF:scalar:L1}). We compare the results for various source distributions, such as Gaussian, Laplace, exponential, and uniform, assuming that the source $X \sim (0,1)$, i.e., zero mean with variance $\var{X} = 1$. In Fig. \ref{fig:RDPF:scalar:L2}, we also compare the estimated result with the SLB derived in Theorem \ref{th:ShannonLowerbound}.
In Fig. \ref{fig:RDPF:scalar:L2}, the Gaussian source case allows us to quantify the algorithm estimation accuracy by comparing it with the $\Rslb$, which in this case represents the exact PR-RDPF. Regarding the other cases, the numerical results show that the bound $\Rslb$ behaves similarly to the SLB of the classical RDF, that is, being tight only in the low distortion (high resolution) regime, while becoming loose at the moderate to high distortion regimes.

\subsubsection*{Vector Case} 
\begin{figure}
    \centering
    \begin{subfigure}{0.49\linewidth}
    	\centering \includegraphics[width= 0.9\linewidth]{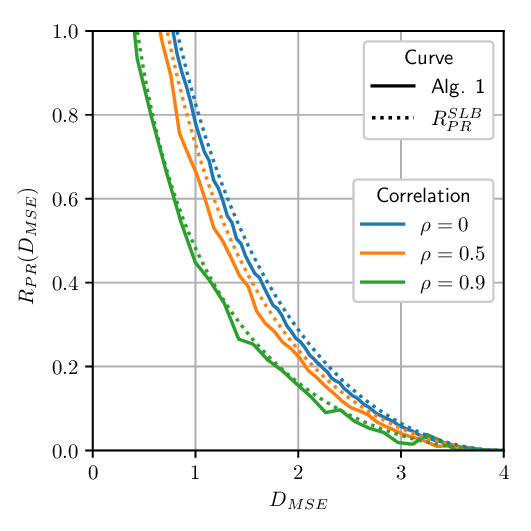}
    	\caption{} \label{fig:RDPF:Vector:Gaussian}
    \end{subfigure}
    \begin{subfigure}{0.49\linewidth}
        \centering \includegraphics[width= 0.9\linewidth]{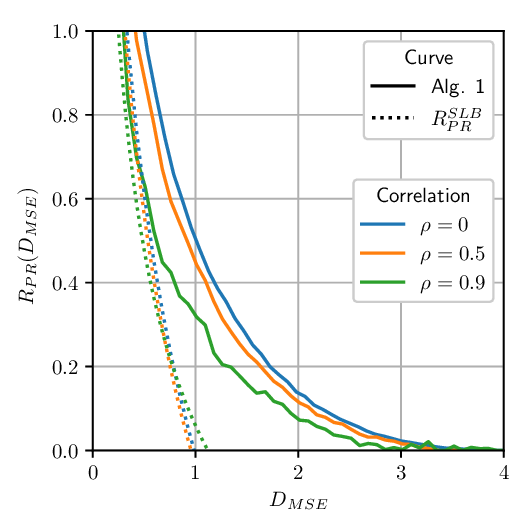}
    	\caption{} \label{fig:RDPF:Vector:Expon}
    \end{subfigure}
    \caption{PR-RDPF under MSE distortion metric for a (a) Gaussian, and (b) exponential bivariate source.} 
\label{fig:Vector:RDPF}
\end{figure}

We estimate the PR-RDPF under an MSE distortion metric for correlated bivariate sources, considering the cases where the source marginals are either Gaussian (see Fig. \ref{fig:RDPF:Vector:Gaussian}) or exponentially (see Fig. \ref{fig:RDPF:Vector:Expon}) distributed with zero mean and variance $\var{} = 1$. In both cases, the multivariate distribution is constructed by imposing a Gaussian coupling\footnote{For more details on parametric copula models, we refer the reader to \cite{durante:2010}.} with variable correlation coefficient $\rho \in \UI$ on the considered marginal distributions. By changing $\rho$, we analyze the cases where the bivariate source presents independent ($\rho$ = 0), mildly correlated ($\rho = 0.5$) and highly correlated ($\rho = 0.9$) marginals.
In Fig. \ref{fig:RDPF:Vector:Gaussian}, we demonstrate a comparison between the Gaussian PR-RDPF estimate obtained via  Alg. $ \ref{alg:MIminimization}$  with the $\Rslb$ obtained in \eqref{eq:shannonLB} with the term $\Rpr^G(D)$ computed via the optimal adaptive reverse-water-filling solution of \cite[Corollary 3]{serra:2023:computation}, which results into a tight $\Rslb(D)$. We observe that Alg. $ \ref{alg:MIminimization}$ provides a very good estimate of the Gaussian PR-RDPF for all the selected $\rho$. We also notice that the estimation error when using Alg. $ \ref{alg:MIminimization}$ remains stable in the low to moderate correlation cases while showing a slightly noisier behavior (fluctuations) in the high correlation case. Contrary to Fig. \ref{fig:RDPF:Vector:Gaussian}, in Fig. \ref{fig:RDPF:Vector:Expon} we observe that beyond high resolution (low distortion), the exponential PR-RDPF estimate obtained via  Alg. $ \ref{alg:MIminimization}$ is much tighter compared to the $\Rslb$. In fact, the latter demonstrates a similar behavior to the SLB of the classical RDF for the multivariate non-Gaussian case. 

\newpage

\IEEEtriggeratref{14}

\bibliographystyle{IEEEtran}
\bibliography{strings, biblio}

\newpage

\appendices

\section{Proof of Theorem \ref{theorem:EquivalenceEOTRD}} \label{proof:EquivalenceEOTRD}
We start by showing that $\hat{\Pi}$ and \FS{$\bar{\Pi}$ in Definitions \ref{problem:OC-RDF} and \ref{problem:EOT}}, define the same set, i.e., there exists a bijection between two sets. Assuming $\pdf{X}$ to be the source distribution in OC-RDF, then for any Markov kernel $\pdf{Y|X}\in\hat{\Pi}$, the joint pdf $\pdf{Y|X}\cdot\pdf{X}$ lies in $\bar{\Pi}$. Conversely, for any joint distribution $\FS{\pdf{X,Y}} \in \bar{\Pi}$ the Markov kernel $\pdf{Y|X} = \frac{\pdf{X,Y}}{\pdf{X}}$ belongs to $\hat{\Pi}$.
Hence, there is a one-to-one mapping between the optimization variables of \FS{Definitions \ref{problem:OC-RDF} and \ref{problem:EOT}}.\\
Let $(D,\lambda)$ be the pair composed by the distortion level in \FS{the constraint of \eqref{opt:RDPF}} and the associated Lagrangian multiplier. Then, the Lagrangian functional of \FS{Definition \ref{problem:OC-RDF}} for distortion level $D$ is defined as
\begin{align}
    \mathcal{L}_{RD}(\pdf{Y|X},\lambda) &\FS{\triangleq} I(X,Y) + \lambda \E[\pdf{Y|X}\pdf{X}]{\Delta(X,Y)}.\label{lagrangian:de_2}
\end{align}
Similarly, the Lagrangian functional associated with \FS{Definition \ref{problem:EOT}} is defined as
\begin{align}
    \mathcal{L}_{EOT}(\pdf{X,Y},\epsilon) &\FS{\triangleq} \E[\pdf{X,Y}]{\Delta(X,Y)} + \epsilon I(X,Y).\label{lagrangian:de_3}
\end{align}
\FS{Based on \eqref{lagrangian:de_2}, \eqref{lagrangian:de_3}}, we observe that the following relation holds
\begin{align*}
    \mathcal{L}_{RD}\left(\frac{\pdf{X,Y}}{\pdf{X}},\lambda\right) = \mathcal{L}_{EOT}\left(\pdf{X,Y},\frac{1}{\lambda}\right) 
\end{align*}
\FS{hence} 
\begin{align}
    \argmin_{\pdf{X,Y} \in \bar{\Pi}} \mathcal{L}_{EOT}\left(\pdf{X,Y},\frac{1}{\lambda}\right) &= \argmin_{\pdf{X,Y} \in \bar{\Pi}} \mathcal{L}_{RD}\left(\frac{\pdf{X,Y}}{\pdf{X}},\lambda\right)\nonumber \\
    & = \pdf{X} \cdot \argmin_{\pdf{Y|X} \in \hat{\Pi}} \mathcal{L}_{RD}(\pdf{Y|X},\lambda).\label{proof:thm:1} 
\end{align}
\FS{As a result, \eqref{proof:thm:1} shows that the solution of Definition \ref{problem:EOT} for $\epsilon = \frac{1}{\lambda}$ is uniquely determined by the solution of Definition \ref{problem:OC-RDF} for the pair $(D,\lambda)$. This completes the proof.}

\section{Proof of Lemma \ref{lemma:MI2KL}} \label{proof:MI2KL}
From the definitions of $I(X,Y)$ and $\E[\cdf{X,Y}]{\Delta(X,Y)}$, \eqref{eq:copula:generalMI} and \eqref{eq:copula:distortion} can be derived as
    \begin{align*}
        &I(X,Y) = \int_{\R^{2d}} \pdf{XY}(\bx,\by) \log \left( \frac{\pdf{XY}(\bx,\by)}{\pdf{X}(\bx)\pdf{Y}(\by)} \right) d\bx d\by\\
        &\stackrel{(a)}{=} \int_{\R^{2d}} \cu{X,Y}(\vcdf{X}(\bx),\vicdf{Y}(\by)) \cdot \\
        & \quad\log \left( \frac{\cu{X,Y}(\vcdf{X}(\bx), \vicdf{Y}(\by))}{\cu{X}(\vcdf{X}(\bx)) \cu{Y}(\vicdf{Y}(\by))}\right) \prod_{i = 1}^d d\cdf{X_i}(x_i)d\cdf{Y_i}(y_i) \\
        & \stackrel{(b)}{=} \int_{\UI^{2d}} \cu{X,Y}(\bu_x,\bu_y) \log \left( \frac{\cu{X,Y}(\bu_x,\bu_y)}{\cu{X}(\bu_x)\cu{Y}(\bu_y)} \right) d\bu_x d\bu_y\\
        & = \KL(\Cu{X,Y}||\Cu{X} \otimes \Cu{Y})
       \end{align*}
    \begin{align*}
    &\E[\cdf{X,Y}]{\Delta(X,Y)} = \int_{\R^{2d}} \Delta(\bx,\by) \pdf{XY}(\bx,\by) d\bx d\by\\
    &\stackrel{(a)}{=} \int_{\R^{2d}} \Delta(\bx,\by) \cu{X,Y}(\vcdf{X}(\bx), \vcdf{Y}(\by)) \prod_{i = 1}^d d\cdf{X_i}(x_i)d\cdf{Y_i}(y_i) \\
    & \stackrel{(b)}{=} \int_{\UI^2} \Delta \left(\vicdf{X}(\bu_x),\vicdf{Y}(\bu_y)\right) \cu{X,Y}(\bu_x,\bu_y)  d\bu_x d\bu_y\\
    & = \E[\cu{X,Y}]{\Delta\left(\vicdf{X}(U_X),\vicdf{Y}(U_Y)\right)}
    \end{align*}
    where (a) \FS{follows} from the application of Corollary \ref{corollary:SklarTheoremForPdf} on $\pdf{X,Y}$,$\pdf{X}$, and $\pdf{Y}$, \FS{and} (b) \FS{follows} from the change of variables $\bu_x = \vcdf{X}(\bx)$ and $\bu_y = \vcdf{Y}(\by)$.

\section{Proof of Theorem \ref{theorem:OptimalProjection}} \label{proof:OptimalProjection}
The proof follows \FS{similar} steps \FS{to} the proof of \cite[Theorem 3.1]{samo:2021:copula} with some technical differences, \FS{hence at certain points we skip the heavy mathematical details for ease of readability}. \FS{In particular,  we project on the product copula d.f. $R$, instead of the $I$-projection of the uniform distribution $U$ on $\UI^{2d}$, which is considered in \cite[Theorem 3.1]{samo:2021:copula}}. 
\par First, we inquire about the existence of the projection.
\paragraph*{\bf Existence and uniqueness} Under the assumption \FS{of our Theorem} that there exist $P \in \mathcal{B}$ with $\FS{\KL(P||R)} < \infty$, if the convex set $\mathcal{B}$ is variation closed, i.e., closed in the topology induced by the total variation distance \cite[Corollary 7.45]{klenke:2013probability}, then there exists a unique $Q$ being the $I$-projection of $R$ on $\mathcal{B}$.  This property of the set $\mathcal{B}$ \FS{can be} proved using \FS{\cite[Lemma B.1]{samo:2021:copula}}.

\paragraph*{\bf Parametric form of the density of the projection} The projection task can be \FS{facilitated} by defining an intermediate projection step onto the set $\mathcal{A}$ \FS{that corresponds to} the set of d.f. on $\UI^d$ satisfying the distortion constraint \eqref{opt:CopulaDistor}. Clearly, \FS{in this case} $\mathcal{A}$ is convex and $\mathcal{B} \subset \mathcal{A}$.
\par \FS{Since the set $\mathcal{A}$ is defined by linear constraints, then following \cite[Theorem 3.1, (Case A)]{Csiszar:1975}},  \FS{we obtain that} $R_{\mathcal{A}}$ is the unique $I$-projection of $R$ onto $\mathcal{A}$ with density
\begin{align}
    \frac{dR_{\mathcal{A}}}{dR}(\bu) = e^{\mu + \theta \Delta(\vicdf{X}(\bu_x),\vicdf{Y}(\bu_y))} 
\end{align}
and, for all $P \in \mathcal{A}$, holds \FS{that} 
\begin{align}
    \KL(P||R) = \KL(P||R_\mathcal{A}) + \KL(R_\mathcal{A}||R).\label{eq:pitagora}
\end{align}
\FS{Moreover}, under the result of \cite[Theorem 2.3]{Csiszar:1975}, if $R$ has $I$-projection $R_{\mathcal{A}}$ on $\mathcal{A}$, and $I$-projection $Q_{\mathcal{B}}$ on $\mathcal{B}$, and if \eqref{eq:pitagora} holds for all $P \in \mathcal{A}$, then $Q_\mathcal{B}$ is the unique $I$-projection of $Q_\mathcal{A}$ onto $\mathcal{B}$. 

\FS{Since the set $\mathcal{B}$ is defined by imposing a constraint on the marginals of the measure, and $Q_\mathcal{A}$ has $I$-projection on $\mathcal{B}$, then using \cite[Theorem 3.1, (Case B)]{Csiszar:1975}}, there exist nonnegative scalar functions $g_i$ with $\log(g_i) \in l_1(\UI)$ for $i=1,\ldots,2d$ such that
\begin{align}
        \frac{dQ}{dR_{\mathcal{A}}}(\bu) = \prod_{i = 1}^{2d} g_{i}(u_i).\nonumber
\end{align}
Therefore, $Q$ has density with respect to $R$ given by  $\frac{dQ}{dR} = \frac{dQ}{dR_{\mathcal{A}}} \frac{dR_{\mathcal{A}}}{dR} = \eqref{eq:copula:optimaldensity}$. This \FS{completes} the proof.

\section{Proof of Theorem \ref{theorem:convergenceRelaxation}} \label{proof:convergenceRelaxation}

\done{Let $\hQ_i$ and $\mathcal{A}_i$ be, respectively, the copula distribution solution of Problem \ref{problem:MomentConstrainedRDPF} and its constraint set, for a number $i$ of constraints on the moments of each marginal. Furthermore, let the constraint set and optimal solution of Problem \ref{problem:ProjectionInfoRDPF} be denoted with $\mathcal{B}$ and $Q^*$, respectively. Our goal is to prove that the sequence $\{\hQ_i\}_{i = 0,1,\ldots}$ convergences to $Q^*$.}

By construction, for all $i = 0,1,\ldots$, \FS{it} holds \FS{that} $\mathcal{B} \subseteq \mathcal{A}_{i+1} \subseteq \mathcal{A}_{i}$. As a consequence of \cite[Theorem 2.3]{Csiszar:1975}, we can characterize $\hQ_{i+1}$ and $Q^*$ as the $I$-projections of $\hQ_i$ onto the sets $\mathcal{A}_{i+1}$ and $\mathcal{B}$, respectively. Then, for all $i = 0,1,\ldots$, the following geometric relation \FS{holds (see \cite[Equation 3.1]{Csiszar:1975})}
\begin{align}
    \KL(\hQ_i||Q^*) = \KL(\hQ_{i}||\hQ_{i+1}) + \KL(\hQ_{i+1}||Q^*). \label{eq:PitagoraTheoremKL}
\end{align} 
Recursively, applying \eqref{eq:PitagoraTheoremKL} $k + 1$ times leads to
\begin{align*}
    \KL(\hQ_i||Q^*) = \sum_{j = i}^{i + k} \KL(\hQ_{j}||\hQ_{j+1}) + \KL(\hQ_{k+1}||Q^*)
\end{align*}
from which we \FS{immediately obtain} 
\begin{align*}
\sum_{j = i}^{i + k } \KL(\hQ_{j}||\hQ_{j+1}) \le \KL(\hQ_i||Q^*). 
\end{align*}
Since \FS{ we assume that $\KL(\hQ_i||Q^*)<\infty$}, then, necessarily $\lim_{k \to \infty} \KL(\hQ_{k}||\hQ_{k+1}) = 0$, implying the convergence of the sequence $\{\hQ_i\}_{i = 0,1,\ldots}$ in KL-divergence. 
\par To prove that the limit of the sequence is $Q^*$, we transform Problem \ref{problem:ProjectionInfoRDPF} into a specific instance of Problem \ref{problem:MomentConstrainedRDPF} under an infinite number of marginals moments constraints. As a consequence of the uniqueness of solutions of the Hausdorff moments problem \cite{shohat:1950problem}, any \FS{RV} on $\UI$ \FS{that respects} $\E{u^n} = \alpha_n$ for all $n \in \N$, is necessarily uniformly distributed. This allows us to transform the uniform marginal constraints in Problem \ref{problem:ProjectionInfoRDPF} into a set of countably infinite marginal constraints. 
\par In such form, the only difference between Problems \ref{problem:ProjectionInfoRDPF} and \ref{problem:MomentConstrainedRDPF} resides in the finite number $i$ of \FS{the} moment constraints of the latter. \FS{Hence}, since $Q^* = \hQ_\infty$, the limit $\lim_{i \to \infty} \KL(\hQ_i||Q^*) = 0$ holds. This \FS{completes} the proof.

\section{Proof Theorem \ref{theorem:ELProjection}} \label{proof:ELProjection}
Let $\mathcal{M}^+(\UI^{2d})$ \FS{denote} the set of measurable functions on $\UI^{2d}$. We notice that the constraint set $\mathcal{A}$ of Problem \ref{problem:MomentConstrainedRDPF} is defined by linear constraints in the copula d.f. $C_{X,Y}$. The results of \cite[Theorem 3.1, (Case A)]{Csiszar:1975} ensure that a unique projection $Q$ of $R$ on $\mathcal{A}$ exists with $\KL(Q||R) < \infty$, hence $\frac{dQ}{dR} \in \mathcal{M}^+(\UI^{2d})$. 
More generally, any function $g \in \mathcal{M}^+(\UI^{2d})$ defines a probability measure $dG = g dR$ on $\UI^{2d}$ under the condition that $\int_{\UI^{2d}} g(\bu) dR(\bu) = 1$. This \FS{enables the definition of} an optimization problem over the set $\mathcal{M}^+(\UI^{2d})$ equivalent to Problem \ref{problem:MomentConstrainedRDPF} as follows
\begin{align}
&\min_{g \in \mathcal{M}^+(\UI^{2d})}\int_{\UI^{2d}}  \log\left(  g(\bu) \right) g(\bu) dR(\bu). \\
\textrm{s.t.}
& \int_{\UI^{2d}} g(\bu) dR(\bu) = 1 \label{eq:EL:Norm}\\
& \int_{\UI^{2d}} \Delta(\vicdf{X}(\bu_x),\vicdf{Y}(\bu_y)) g(\bu) dR(\bu) = D \label{eq:EL:Rec}\\
& \int_{\UI^{2d}} u_{i}^n g(\bu) dR(\bu) = \alpha_n \qquad (i,n) \in I \label{eq:EL:Uni}
\end{align}
Indicating with $\mu', \theta, \{ \nu_{i,n} \}$ the Lagrangian multipliers associated with constraints \eqref{eq:EL:Norm}-\eqref{eq:EL:Uni}, we define the Lagrangian functional of the problem as 
\begin{align}
\begin{split}
L(g, \mu', \theta, \{ \nu_{i,n} \}) &= \int S(g(\bu), \mu', \theta, \{ \nu_{i,n} \}) dR(\bu) \\
& \quad + V( \mu', \theta, \{ \nu_{i,n} \})
\end{split} \label{eq:EL:Lagrangian} \\
\begin{split}
S(z, \mu', \theta, \{ \nu_{i,n} \}) &= z\Big[\log(z) - \mu' - \sum_{(i,n)\in I} \nu_{i,n} u_{i}^n \\
& \quad \qquad -\theta\Delta(\vicdf{X}(\bu_x),\vicdf{Y}(\bu_y)) \Big]
\end{split} \nonumber \\
V(\mu', \theta, \{ \nu_{i,n} \}) &= \mu' + \theta D +\sum_{(i,n) \in I}  \nu_{i,n} \alpha_n. \nonumber
\end{align}
\FS{By applying} the Euler-Lagrange equation \cite{gelfand:2000calculus}, \FS{we} characterize necessary conditions for the function $g$ to be an extreme point for \eqref{eq:EL:Lagrangian}. If $g^*$ is an extreme point \done{ of \eqref{eq:EL:Lagrangian} }, then the necessary {\it stationarity condition} holds, \FS{i.e.,}
\begin{align}
\frac{dS}{dz}(g^*) = 0 \nonumber
\end{align}
from which we \FS{obtain}
\begin{align}
\begin{split}
g^*(\bu) = \frac{dQ}{dR}(\bu) &= \exp\Big[(\mu' - 1)  + \sum_{(i,n)\in I} \nu_{i,n} u_{i}^n \\
& \qquad + \theta\Delta(\vicdf{X}(\bu_x),\vicdf{Y}(\bu_y))\Big],
\end{split}
\end{align}
which is equivalent to \eqref{eq:copula:ELdensity} by considering $\mu = \mu' -1$.
\par To determine the optimal values of the Lagrangian  \done{multipliers $\mu', \theta, \{ \nu_{i,n} \}$}, we leverage Lagrangian duality \FS{theorem \cite{rockafellar1970convex}} and define the dual problem as
    \begin{align*}
        &\max_{(\mu', \theta, \{ \nu_{i,n} \}_{(i,n)\in I} )} L(g^*, \mu', \theta, \{ \nu_{i,n} \}_{(i,n)\in I}) \\
        & = \min_{(\mu, \theta, \{ \nu_{i,n} \}_{(i,n)\in I} )} - L(g^*, \mu + 1, \theta, \{ \nu_{i,n} \}_{(i,n)\in I}) \\
        & = \eqref{eq:copula:ELdual}. 
    \end{align*}
\FS{This concludes the proof.}

\section{Proof Lemma \ref{lemma:StrictCovexity}} \label{proof:StrictConvexity}
\FS{The proof makes use of the Hessian matrix of the optimization problem in \eqref{eq:copula:ELdual}.}   \FS{Specifically, define} the vector $\mathbf{l} \triangleq ((\mu, \theta, \{ \nu_{i,n} \}_{(i,n)\in I})) \in \R^{2 + 2d \cdot N}$ and the mapping $\mathbf{\omega}(\bu)\triangleq (1, \Delta(\vicdf{X}(\bu_x),\vicdf{X}(\bu_y)), u_1,\ldots,u_1^N,\ldots,u_{2d}^N)$. \FS{Then,} the Hessian of the optimization problem \eqref{eq:copula:ELdual} can be expressed as
\begin{align*}
\int_{I^{2d}} \mathbf{\omega}(\bu)\mathbf{\omega}(\bu)^T dQ = \E[Q]{\mathbf{\omega}(\bu)\mathbf{\omega}(\bu)^T},
\end{align*}
which, due to the linear independence of the components of $\mathbf{\omega}(\bu)$, is a strictly positive definite matrix. This is a sufficient condition for the strict convexity of \eqref{eq:copula:ELdual}. \FS{This completes the proof}.

\section{Proof of Theorem \ref{th:ShannonLowerbound}} \label{proof:ShannonLowerbound}
We start by considering the scalar version of the proposed problem, i.e., $\mathcal{S} \triangleq \{\pdf{X}: \E[\pdf{X}]{(X - \E{X})^2} \le \var{} \}$. 
\FS{Define} the constraint set $\mathcal{H}(\pdf{X}, D)$ \FS{as follows}
\begin{align*}
\mathcal{H}(\pdf{X}, D) = \{ \pdf{Y|X}: \E[\pdf{X}\pdf{Y|X}]{||X - Y||^2} \le D, X \sim Y \}.
\end{align*}
\FS{Then, by definition of the PR-RDPF}, we obtain
    \begin{align}
        \Rpr(D) &= h(X) - \max_{ \pdf{Y|X} \in \mathcal{H}(\pdf{X}, D)} h(X|Y) \nonumber \\
        & \stackrel{(a)}{=} h(X) - \max_{ \pdf{Y|X} \in \mathcal{H}(\pdf{X}, D)} h(Y|X) \nonumber \\
        & \stackrel{(b)}{\ge} h(X) - \max_{\pdf{Z} \in \mathcal{S}}\max_{ \pdf{Y|Z} \in \mathcal{H}(\pdf{Z}, D)} h(Y|Z) \label{eq:shannonLB:eq1}
    \end{align}
where (a) \FS{follows} by observing that $h(X|Y) = h(Y|X)$ under the constraint $X \sim Y$, and (b) \FS{follows} from the maximization over the set of sources $\mathcal{S}$.
\par Consider the joint distribution on $(Z,Y)$ with $Z \sim Y$. \FS{Then, the conditional variance of the RV $Y$ conditioned on $Z$} can be expressed as $\var{Y|Z} = \var{Z}(1 - \rho^2)$, hence the
constraint set $\mathcal{H}(\pdf{Z}, D)$ can be simplified \FS{to}
\begin{align}
    \E[\pdf{Z}\pdf{Y|Z}]{||Z-Y||^2} = 2\var{Z}(1 - \rho) \le D \nonumber \\
    \implies \var{Y|Z} \le \var{Z}\left(1 - \left(1 - \frac{D}{\var{Z}}\right)^2\right) = D'. \label{eq:shannonLB:eq4}
\end{align}
\par Since \eqref{eq:shannonLB:eq4} constraints only the second moment of the distribution $\pdf{Y|Z}$, we can infer that the maximum $h(Y|Z)$ is attained by $\pdf{Y|Z} \sim N(0, D')$. 
\done{Furthermore, since \eqref{eq:shannonLB:eq4} depends only on the second moment of the source $\var{Z}$, the maximization over the set of source distributions has to satisfy only the distribution constraint $Z \sim Y$. Assuming $Y|Z$ to be Gaussian, we can select $Z$ to be also Gaussian distributed with $\pdf{Z} \sim N(0,\var{})$, which ensures that $Y$ will have the same distribution.}
Therefore, the distribution on $(Y,Z)$ maximizing  $h(Y|Z)$ is itself Gaussian and coincides with the \FS{PR-RDPF} achieving distribution assuming a Gaussian source $X^* \sim N(0,\var{})$. This allows \FS{the characterization of} the following equality
\begin{align}
    \max_{\pdf{Z} \in \mathcal{S}}\max_{ \pdf{Y|Z} \in \mathcal{H}(\pdf{Z}, D)} h(Y|Z) &= \max_{ \pdf{Y|X^*} \in \mathcal{H}(\pdf{X^*}, D)} h(Y|Z) \\ 
    & =\todo{\Rpr^G(D)} - h(X^*) \label{eq:shannonLB:eq2}
\end{align}
which, together with \eqref{eq:shannonLB:eq1} \FS{yields} \eqref{eq:shannonLB}. 
\par For the general vector case, we consider that for every source $Z \sim \pdf{Z} \in \mathcal{S}$, \FS{we have marginals} $\{Z_i\}_{i = 1,\ldots, N}$ \FS{with} variance $\var{Z_i} = \lambda_{i}$, where $\{\lambda_{i}\}_{i = 1,\ldots, N}$ is the set of eigenvalues of $\Sigma$. Then, from \eqref{eq:shannonLB:eq1} we \FS{obtain}
\begin{align}
    &\max_{\pdf{Z} \in \mathcal{S}}\max_{ \pdf{Y|Z} \in \mathcal{H}(\pdf{Z}, D)} h(Y|Z) \nonumber \\
    & \stackrel{(a)}{\le} \max_{\pdf{Z} \in \mathcal{S}}\max_{ \pdf{Y|Z} \in \mathcal{H}(\pdf{Z}, D)} \sum_{i = 1}^N h(Y_i|Z_i) \nonumber\\
    & \stackrel{(b)}{\le}\max_{\substack{D_i : \sum_i^N D_i = D }} \max_{\substack{\pdf{Z} \in \mathcal{S}, \pdf{Y|Z} \\ \E{||Z_i - Y_i||^2} \le D_i ~ \forall i = 1, \ldots, N\\ Z \sim Y }} \sum_{i = 1}^N h(Y_i|Z_i)  \nonumber \\
    & \stackrel{(c)}{\le} \max_{\substack{D_i : \sum_i^N D_i = D }}  \sum_{i = 1}^N \max_{\substack{\pdf{Z_i} \in \mathcal{S}, \pdf{Y_i|Z_i} \\ \E{||Z_i - Y_i||^2} \le D_i ~ \forall i = 1, \ldots, N\\ Z_i \sim Y_i }} h(Y_i|Z_i) \nonumber \\
    &  \stackrel{(d)}{\le} \max_{\substack{D_i : \sum_i^N D_i = D }}  \sum_{i = 1}^N \Rpr^{G,i}(D_i) - h(X^*_i)  \nonumber \\
    & = - h(X^*) + \max_{\substack{D_i : \sum_i^N D_i = D }}  \sum_{i = 1}^N \Rpr^{G,i}(D_i) \label{eq:shannonLB:eq3}
\end{align}
where (a) \FS{follows from} the property of the differential entropy $h(Y_i|Z_k, Z_j) \le h(Y_i,Z_k)$; (b) \FS{follows} from the tensorization properties of the MSE distortion; (c) \FS{follows} from breaking the maximization of the sum of functions into the sum of the maximum of each function; (d) \FS{follows} from \eqref{eq:shannonLB:eq2}, by considering $X^* \sim N(0,\Sigma)$ with marginals $X_i^* \sim N(0, \lambda_{i})$ and $\Rpr^{G,i}(D_i)$ being the PR-RDPF for source $X_i$ and distortion level $D_i$. 
Using \cite[Corollary 3]{serra:2023:computation}, we see that the second term of \eqref{eq:shannonLB:eq3} can be shown to be equivalent to the PR-RDPF for the source $X^*$, therefore showing that \eqref{eq:shannonLB:eq1}, and consequently \eqref{eq:shannonLB}, also hold in the general vector case. This \FS{completes} the proof.

\end{document}